\documentclass[pdflatex,sn-mathphys-num]{sn-jnl}
\usepackage{graphicx}%
\usepackage{multirow}%
\usepackage{amsmath,amssymb,amsfonts}%
\usepackage{amsthm}%
\usepackage{mathrsfs}%
\usepackage[title]{appendix}%
\usepackage{xcolor}%
\usepackage{textcomp}%
\usepackage{manyfoot}%
\usepackage{booktabs}%
\usepackage{algorithm}%
\usepackage{algorithmicx}%
\usepackage{algpseudocode}%
\usepackage{listings}%
\usepackage{comment}

\theoremstyle{thmstyleone}%
\theoremstyle{thmstyletwo}%

\theoremstyle{thmstylethree}%

\begin{document}

\title[The Birth of Gravitational Particle Creation: the Enduring Legacy of Leonard Parker's 1966 Thesis]{The Birth of Gravitational Particle Creation: the Enduring Legacy of Leonard Parker's 1966 Thesis}

\author[1]{\fnm{Antonio} \sur{Ferreiro}}\email{a.e.ferreirodeaguiar@uu.nl}
\equalcont{These authors contributed equally to this work.}

\author[2]{\fnm{Jos\'e} \sur{Navarro-Salas}}\email{jnavarro@ific.uv.es}
\equalcont{These authors contributed equally to this work.}

\author[3]{\fnm{Silvia} \sur{Pla}}\email{silvia.pla-garcia@tum.de}
\equalcont{These authors contributed equally to this work.}

\affil[1]{\orgdiv{Department of Mathematics}, \orgname{Utrecht University}, \orgaddress{\street{Princetonplein 5}, \city{Utrecht}, \postcode{3584 CC}, \state{State}, \country{The Netherlands}}}

\affil[2]{\orgdiv{Departamento de F\'isica Te\'orica}, \orgname{Instituto de Física Corpuscular (IFIC), CSIC‐Universitat de València, Spain}, \orgaddress{\street{C/ Doctor Moliner 50}, \city{Burjassot}, \postcode{46100}, \state{State}, \country{Spain}}}


\affil[3]{\orgdiv{Physik-Department}, \orgname{Technische Universität München}, \orgaddress{\street{James-Franck-Str.}, \city{Garching}, \postcode{85748}, \country{Germany}}}


\abstract{This paper offers a historical overview of the origins and enduring significance of gravitational particle creation, a groundbreaking discovery first formulated in Leonard Parker's 1966 doctoral thesis at Harvard University. By tracing the context in which Parker developed this idea and examining its subsequent influence, the paper highlights how the concept of gravitational particle creation advanced the study of quantum field theory in curved spacetime and profoundly shaped modern cosmology, as well as the quantum theory of black holes.}


\maketitle

\section{Introduction}

Leonard Parker's 1966 Harvard University thesis, {\it The Creation of Particles in an Expanding Universe}, opened the door to the study of gravitational particle creation. This phenomenon has since become a cornerstone of modern cosmology, black hole physics, and many of their subsequent developments. Over the decades, generations of physicists have learned about particle production by gravitational fields, yet the pioneering work that first revealed and explained this effect often remains underappreciated. 

To address this, we have recently retyped and reedited Parker's original thesis to make it accessible through the arXiv \cite{ParkerThesis}. In this work, our goal is to provide a concise account of the historical context surrounding Parker's breakthrough and to discuss its key physical implications from a contemporary perspective. 
To capture the spirit of that moment, it is worth quoting Paul Davies \cite{MJC07}, who reflected on Parker's thesis and his subsequent publications in Physical Review Letters \cite{Parker68} and Physical  Review \cite{Parker69, Parker71}:
\begin{quote}

{\it It was a leap in the dark, and it turned out to be exactly what was needed. It took a lot of courage to embark on that. He could not have anticipated how important it would all turn out to be 10 or 15 years from then. He set the train in motion for a decade or more of similar work.}
\end{quote}
This new field, now known as Quantum Field Theory in Curved Spacetime, emerged in the 1960s with Parker's pioneering thesis and reached maturity in the 1980s, as exemplified by the influential monograph by Birrell and Davies \cite{Birrell-Davies}. 

Our account is organized as follows. In Section 2, we provide a brief outline of the historical background leading up to and around 1962, the year Parker began his doctoral research. Section 3 examines the period 1962–1966, during which his thesis was developed. We summarize the key physical ideas underlying the discovery of gravitational particle creation, emphasizing Parker's seminal frequency-mixing mechanism. We do not delve into the mathematical details—rigorous though they are—but rather focus on the central physical insights that guided this pioneering work. In Section 4, we describe how the results of Parker's thesis disseminated through Princeton, Moscow, and Cambridge, stimulating a wave of new developments, including the Fulling–Davies–Unruh effect, the damping of anisotropic cosmic expansions, and black hole radiation. These topics are examined in greater detail in Section 5. 
Finally, Section 6 presents our concluding reflections on the lasting influence of Parker's thesis—an impact that motivated our effort to retype and edit the original work \cite{ParkerThesis}.
  
\section{Before 1963}

It has now been a century since the foundations of quantum mechanics were laid, largely through the independent yet complementary breakthroughs of Heisenberg and Schrödinger, themselves building upon the pioneering works of Planck, Einstein, Bohr, de Broglie, and others. Dirac subsequently extended the framework of quantum mechanics by uniting it with special relativity, thereby formulating relativistic quantum theory and predicting the existence of antimatter. Following Dirac’s theoretical discovery of the positron, Heisenberg and his student Euler, in a landmark 1936 paper \cite{Heisenberg-Euler} uncovered the instability of the Dirac sea—a heuristic precursor to the modern concept of the quantum vacuum.

Dirac's relativistic theory of the electron and positron, though groundbreaking, proved to be incomplete. It failed to align with the experimental results announced at the Shelter Island Conference in 1947, casting doubt on the prevailing framework of quantum electrodynamics. This challenge was soon met by Julian Schwinger,  who successfully explained the anomalous magnetic moment of the electron. During this same period, quantum field theory, or more precisely quantum electrodynamics, reached full maturity through the independent but ultimately equivalent formulations developed by Richard Feynman, Sin-Itiro Tomonaga, and Julian Schwinger.

A decisive step forward came in the early 1950s, when Schwinger, using the powerful methods of quantum field theory that he had pioneered, delivered the first precise prediction of particle–antiparticle pair creation in strong electromagnetic fields \cite{Schwinger51}.
Due to the minimal-coupling interaction between a quantized  Dirac field and a prescribed classical electromagnetic field $A_\mu$
\begin{equation} -e \bar \Psi \gamma^\mu \Psi A_\mu
\end{equation}
the quantum vacuum turns out to be unstable. Following Schwinger \cite{Schwinger51}

\begin{quote}

{\it We shall now simply remark that, to extend our results to pair-producing fields, it is merely necessary to $(\cdots)$  interpret the positive imaginary contribution to $W$ thus obtained with the statement that
$$
\left|e^{i W}\right|^2=e^{-2 \operatorname{Im} W}
$$
represents the probability that no actual pair creation occurs throughout the field's history. } 
\end{quote}
$W$ is the so-called effective action in the presence of the external electromagnetic field.\footnote{The expression $\left|e^{i W}\right|^2$ can be translated to the usual Dyson's formulaes in the interaction picture $\left|e^{i W}\right|^2 = |\langle 0 |S|0\rangle|^2 =|T\exp{[-ie\int d^4x \bar \Psi \gamma^\mu \Psi A_\mu]}|^2 $. $T$ is the time-ordering operation, $|0\rangle$ is the vacuum of the Dirac field in the interacting picture, and $S$ is the scattering operator.} It can be expressed in terms of functional determinants and the proper-time formalism. Evaluated for constant fields, led to the Heisenberg-Euler effective action \cite{Heisenberg-Euler}. The probability that the vacuum remains stable is $ e^{-2 \text{Im} W} $. The imaginary part of $W$ appears when an electric field is present ($E\neq 0$), and Schwinger finds, for a constant field: 
\begin{equation}\label{Schwingerpairrate}
2 \text{Im} W = VT \frac{(eE)^2}{4\pi^3c\hbar^2}\sum_{n=1}^\infty \frac{1}{n} e^{-\frac{n\pi m^2c^3}{\hbar|eE|}}
\end{equation}
where $VT$ is the spacetime volume. This is the famous Schwinger rate for the electron-positron pair creation \cite{Schwinger51}.  Remarkably, the result cannot be expressed as a (perturbative) series expansion in powers of the coupling constant. 

 The success of the renormalization program of perturbative quantum electrodynamics—namely, the absorption of ultraviolet divergences arising in perturbation theory into redefined parameters of the theory—came to be regarded as a natural criterion for selecting viable field-theoretic models of the fundamental interactions. Nevertheless, despite the early achievements of quantum field theory, no renormalizable model for the electroweak or strong interactions was initially available.
As a result, confidence in the idea of the quantized field as a primary entity gradually declined, with notable exceptions, in many academic centers from the late 1950s through the late 1960s.\footnote{Among the principal alternatives were current algebra and $S$-matrix theory. In the early 1960s, J. Schwinger and his group at Harvard remained  one of the most influential defenders of QFT as a fundamental framework.
} For a historical overview, see for instance \cite{Gross92}.

On the gravity and general relativity front, progress was slow, and developments remained limited for several decades. In the mid-1950s, the editor-in-chief of the prestigious Physical Review even drafted an editorial announcing that the journal would no longer accept articles on gravitation. Although this policy was never actually implemented, the draft stirred considerable reactions within the community. In part as a response, the 1957 conference {\it On the Role of Gravitation in Physics}\footnote{Recent transcripts of this conference [C. M. DeWitt and D. Rickles, eds., The Role of Gravitation in Physics. Report from the 1957 Chapel Hill Conference (Max Planck Research Library for the History and Development of Knowledge, 2011)]. This conference marked the first in the GRn series, which is now held under the auspices of the International Society on General Relativity and Gravitation.} was organized in Chapel Hill, North Carolina. Led by Cécile DeWitt-Morette and Bryce DeWitt, the event brought together prominent physicists including John Wheeler and Richard Feynman.

Starting in the 1960s, however, general relativity entered what has been called \textit{its renaissance period} (as described by Clifford Will  \cite{Will1986,  Will2018}), with rapid advances both on the experimental and theoretical fronts. Experimentally, the theory began to receive robust empirical support, notably through the Pound–Rebka experiment (1960) which confirmed gravitational redshift, and the observation of the Shapiro time delay (1964). On the theoretical side, decisive progress was made progressively in proving the existence and properties of gravitational waves (one of the main topics of discussion of the 1957 Chapel Hill conference)  and black holes,  using robust and increasingly abstract mathematical techniques. The work by Bondi, Dicke, Sciama, Penrose, and others fixed the path to use these mathematical tools to obtain reliable theoretical results —results that would eventually be probed fifty years later with the detection of gravitational waves \cite{LIGO15} and the imaging of black holes \cite{Imaging}.

The other topic discussed at the 1957 conference was the need for a quantized theory of gravity. The answer of Feynman and others was affirmative. At that time, however, no one knew how to tackle the problem—not even by applying the (perturbative) methods that had proven so successful in quantum electrodynamics. Feynman pursued the issue for several years and, at the GR3 conference held in Warsaw in 1962, argued that the theory required an additional element: fictitious fields —now referred to as ghost fields—to ensure unitarity at the one-loop level in perturbation theory \cite{Feynman63}. In 1964, DeWitt extended Feynman’s work by introducing ghost fields at the two-loop level \cite{DeWitt64}.\footnote{In the same year, 1964, DeWitt published his celebrated Les Houches lectures —an influential and comprehensive work later expanded into a book \cite{DeWitt64b}. It presents a global approach to quantum field theory, primarily formulated in terms of covariant Green’s functions and extending Schwinger's methods to curved spacetime.} Two years later, he developed a comprehensive framework for using ghost fields to quantize both gravitational and non-abelian gauge fields to all orders \cite{DeWitt67b, DeWitt67c}. However, this work was not published until 1967, appearing only weeks before the famous paper by Faddeev and Popov \cite{Faddeev-Popov67}.\footnote{This framework was subsequently employed by G. ’t Hooft and M. Veltman to demonstrate the renormalizability of the Glashow–Weinberg–Salam electroweak theory, which marked a significant revival of quantum field theory. Furthermore, it also provided evidence for the non-renormalizability of general relativity in the presence of matter fields \cite{Hooft-Veltman73}.} Parallel to this program, the canonical quantization approach tried to extend Dirac's program of quantization to General Relativity as a constrained Hamiltonian system. The works of Bergmann, Dirac, Arnowit, Deser, Misner, DeWitt, Wheeler, and others provided important steps in this path \cite{Witten62}, but faced critical technical problems.\footnote{Decades later, those problems provided the seeds for the introduction of new canonical variables \cite{Ashtekar87, Barbero94} and  the ``loop representation of quantum general relativity'' \cite{Rovelli-Smolin}.} 
In contrast to the relative coherence of quantum field theory in particle physics and of the general relativity research program, the field of quantum gravity experienced deep transformations and a broad diversification of research directions.

\cite{Bambi:2023jiz}. As we will see in Sections \ref{section5} and \ref{section6}, Parker’s pioneering work laid the groundwork for the theory of quantum fields in curved spacetime, a framework at the intersection of quantum theory and gravity,  and one that will have a fundamental role in the turns taken by the quantum gravity program.

\section{Parker's thesis (1962-1966)}

In this context—marked by the major experimental and theoretical developments of particle physics, the gradual renaissance of general relativity, and the dominant focus on attempts to quantize gravity directly—it was quite striking that Leonard Parker chose to follow a relatively solitary and unconventional path. He decided to devote his  PhD years 
to the study of field quantization in a classical gravitational background.

When Parker began his graduate studies at Harvard University in 1962, he found himself among leading figures in quantum field theory, such as Wendell Furry, Roy Glauber, Sidney Coleman, Sheldon Glashow, and Julian Schwinger. It is therefore unsurprising that he was initially inclined to pursue a PhD in this area. At the same time, the conceptual difficulties surrounding the direct quantization of the gravitational field prompted him to consider a different line of inquiry. As he later recalled in \cite{Parker17}:

\begin{quote}

{\it I wanted to find new consequences of the quantum field
theory of elementary particles in the context of Einstein’s
theory of general relativity. At the time, I felt that quantizing the nonlinear gravitational field itself was so difficult that I would not be able to make significant progress in trying to go beyond the deep work that had already been done in that area. Nevertheless, I felt that it would
be valuable to study quantized elementary particle fields in the curved space-times that were solutions of the non-linear Einstein gravitational field equations.  Luckily, Sidney Coleman agreed to be my thesis advisor on such a project, which was outside the main stream of the time.

I started by looking for new consequences of quantum field theory in the isotropically expanding cosmological space-times that were solutions of the nonlinear equations of general relativity.} 
\end{quote}

In short, the main discovery of this project is that the (isotropic) cosmic expansion of the universe, typically described  by the spatially flat  FLRW metric\footnote{$R(t)$ represents the physical distance at time 
$t$ between two points that are one unit apart in comoving coordinates.}
\begin{equation}\label{metric}
ds^2 = -dt^2 + R(t)^2 (dx^2 + dy^2 + dz^2) 
\ , \end{equation}
implies that the familiar creation  operators of quantum fields evolve into a superposition of both creation and annihilation operators. The simplest field model to consider was a  neutral scalar field $\varphi$, with action functional 
\begin{equation}\label{action}
S= -\frac{1}{2}\int d^4x \sqrt{-g}[g^{\mu\nu}\partial_\mu \varphi \partial_\mu \varphi + m^2 \varphi^2] \ , 
\end{equation}
propagating in the time-dependent metric (\ref{metric}). 
In this case, 
the late time annihilation operator $B_{\vec{k}}$ of comoving momentum ${\vec k}$ is a linear combination of the initial annihilation and creation
operators $A_{\vec{k}}$ and $A_{-\vec{k}}{ }^{\dagger}$,
\begin{equation}\label{Bogo}
B_{\vec{k}}=\alpha_k A_{\vec{k}}+\beta_k A_{-\vec{k}}^{\dagger} \ ,  
\end{equation}
where the complex coefficients $\alpha_k$ and $\beta_k$ must satisfy \begin{equation}|\alpha_k|^2- |\beta_k|^2=1 \ . \end{equation} This appears as a consequence of the mixing of the positive and negative  frequency modes of the quantized field under the influence of the time-dependent gravitational field.\footnote{In other words, within the idealized framework of an asymptotically Minkowskian bounded spacetime, two distinct quantizations arise: one defined with respect to the inertial time $t_{in}$ 
at early times, and another with respect to the inertial time $t_{out}$
at late times. The connection between these two quantizations is given by (\ref{Bogo}). This further made evident the ambiguity associated with the quantum vacuum in a gravitational field.} The transformation (\ref{Bogo}) implies the creation of particle-antiparticle pairs from the vacuum. This can be  seen by evaluating the expectation value of the number of particles created by the expanding universe from the early-time vacuum state  ${\vec k}$
\begin{equation}\label{number}
\langle 0| B_{\vec{k}}^{\dagger} B_{\vec{k}}|0\rangle = \left|\beta_k(t)\right|^2
\ . \end{equation}
Here, $|0\rangle$ is the state annihilated by the operators $A_{\vec{k}}$ for all modes $\vec{k}$.
It is assumed the Heisenberg picture to describe the time evolution of the quantized field, so the state vector does not change with time. The linear transformation (\ref{Bogo}) is an example of a Bogoliubov transformation.\footnote{In the first version of the thesis, the gravitationally induced transformation (\ref{Bogo}) was not identified as a Bogoliubov transformation, originally introduced in a different physical  context, as the author was unaware of this connection.  A comment was added at the end of Appendix AI in the final version of the thesis acknowledging that (\ref{Bogo}) is, in fact, a Bogoliubov-type transformation. Bogoliubov transformations were originally introduced as a means of diagonalizing Hamiltonians in the theories of superfluidity and superconductivity \cite{Bogoliubov1, Bogoliubov2}.}

The expectation value (\ref{number}) cannot be expressed as a power series, and perturbative techniques based on such an expansion are therefore inapplicable—similar to the situation studied by Schwinger in a different context (see (\ref{Schwingerpairrate})).
Furthermore, the observable particle number (\ref{number}), which is a functional of 
$R(t)$ and depends sensitively on the early stages of the expansion, is analyzed in detail during the period when the expansion is still taking place.

The discovery that an expanding universe can spontaneously generate particles was both profound and unexpected. According to the author’s own account \cite{Parker12}, this breakthrough dates back to 1962–1963, and was presented in Chapter 2 of the thesis \cite{ParkerThesis}. The same chapter included  a detailed description of the early-times vacuum state in terms of late-time particle content. It was interpreted as stating that the creation of excitations occurs in pairs with net momentum zero. The probability of finding one pair is totally independent of the probability of finding another pair. The relative probability of observing a pair of excitations is\footnote{For large $k$, the typical behavior is $\left|\frac{\beta_k}{\alpha_k}\right|^2 \sim e^{-ck}$.}
\begin{equation}\label{beta/alpha}
\left|\frac{\beta_k}{\alpha_k}\right|^2
\ . \end{equation}
Equivalently, one may express the early-time vacuum in terms of the late-time excitations as a two-mode squeezed state, in the terminology of modern quantum optics.\footnote{In the context of pure electrodynamics, an interaction of the form $A_\mu J^\mu$, where $J^\mu$ denotes a prescribed classical source and $A_\mu$ the quantized Maxwell  field, was identified by Glauber \cite{Glauber63} as giving rise to a coherent state, thereby laying the foundations of quantum optics. Similarly, in the gravitational setting, expressing the initial vacuum in terms of late-time excitations anticipated the notion of the squeezed state, introduced into quantum optics several years later.} Furthermore, Parker’s thesis also predicted a  stimulated contribution to boson production: if the initial quantum state already contains particles, the creation of additional particles is enhanced in the initially populated modes.

Between 1964 and 1965, Parker extended his analysis to spin-$\tfrac{1}{2}$ fields (Chapter~3), demonstrating, in particular, that massless fields undergo no particle production in an expanding universe. More broadly, he established that free fields of arbitrary integer or half-integer spin satisfying the conformally invariant field equations in a spatially flat FLRW spacetime likewise exhibit no particle creation (Chapter 4). This result arises from conformal invariance: $g_{\mu\nu}(x) \to \Omega^2(x)\  g_{\mu\nu} (x)$. 
It was therefore clearly established that photons, massless spin-$\tfrac{1}{2}$ particles, and spin-$0$ particles with conformal coupling $\xi R \phi$ (with $\xi = 1/6$) cannot be spontaneously created in an isotropically expanding universe.\footnote{The special coupling $\xi = 1/6$ was also identified by Penrose from a purely geometrical perspective \cite{Penrose64b}.}  In sharp contrast, massless quanta obeying the minimally coupled scalar field equations \emph{can} be created—a case that, as was shown several years later, proves to be of considerable physical significance.

It is worth emphasizing again that the methods developed by Parker to predict particle creation differ  from those employed by Schwinger in deriving (\ref{Schwingerpairrate}). Nonetheless, Parker’s framework for gravitational particle creation admits a natural extension to electrodynamics in flat spacetime, wherein the result (\ref{Schwingerpairrate}) emerges in the appropriate limit of a constant electric field.

The absence of spontaneous photon creation in an isotropically expanding universe explains why Schrödinger need not have been alarmed by the “stimulated” process of light he analyzed  in  1939 \cite{Schrodinger39}. This process is often cited as an early anticipation of cosmological particle creation, so let us briefly clarify the situation.\footnote{Parker was unaware of Schrödinger’s work until the early 1970s, when a former colleague of Schrödinger informed him about it \cite{Parker72}.}
In \cite{Schrodinger39}, Schrödinger studied wave packet propagation in an isotropically expanding universe. Although his primary interest lay in electromagnetic waves, he also analyzed  a minimally coupled spin-$0$ wave equation to  model material particles. 
He argued that as the universe expands, a wave packet increases in strength while also creating a weaker wave moving in the opposite direction. Schrödinger saw this effect—similar to classical superradiance—as meaning that a particle (or photon) moving one way could cause the creation of a particle–antiparticle pair: one continuing forward and the other moving backward, both with the same frequency as the original particle.
He referred to this as an “alarming” phenomenon: “Waves traveling in one direction can (not) be rigorously separated from those traveling in the opposite direction”.

We now understand, however, that photons in an isotropically expanding universe satisfy conformally invariant field equations. As a result, the expansion cannot produce photon pairs—neither spontaneously from the vacuum nor through the stimulated mechanism Schrödinger envisioned in \cite{Schrodinger39}.\footnote{Schrödinger later revised his initial conclusions on the behavior of light in a subsequent paper \cite{Schrodinger40}, correcting the views he had presented earlier \cite{Schrodinger39}.} By contrast, Schrödinger’s conclusions were correct for the case of minimally coupled spin-$0$ scalar fields. Nevertheless, his paper remained largely unknown for decades, partly because Schrödinger employed the methods of wave mechanics rather than the formalism of quantum field theory. A complete and rigorous treatment of particle creation—both spontaneous and stimulated—within the framework of quantum field theory had to await the Ph.D. thesis of Leonard Parker.

To conclude this section, it is worth briefly describing another significant development that occurred in parallel with Parker’s thesis, helping to place it in a clearer historical context:

i) As already noted, Bryce DeWitt was deeply engaged in his monumental “trilogy,” \cite{DeWitt67a, DeWitt67b, DeWitt67c} devoted in particular to formulating a consistent perturbation theory for both general relativity and Yang–Mills theory, which required, among other elements, the introduction of ghost fields. Parker’s thesis pursued a complementary direction: the gravitational field was treated as a classical background, while the matter fields were quantized. This more conservative framework led to the discovery of gravitational particle creation —an effect that had been missed within the more ambitious program of attempting to quantize the gravitational field itself. 
The impact of DeWitt’s work became evident, perhaps unexpectedly, in the eventual acceptance of non-Abelian gauge theories and the establishment of the Standard Model of particle physics, following the seminal contributions of ’t Hooft and Veltman.\footnote{As M. Veltman remarked \cite{Veltman75}: {\it DeWitt in his 1964 Letter and in his subsequent monumental work derived most of the things we know now.}} \\

ii) 
Stephen Hawking undertook his Ph.D. project, Properties of Expanding Universes, at Cambridge University, which he completed between 1962 and 1966. Both Hawking’s and Parker’s theses were connected to the Steady State theory of the universe, which remained the dominant cosmological model in the early 1960s. Parker’s thesis demonstrated that the average rate of matter creation predicted by quantum field theory in an expanding spacetime is many orders of magnitude too small to sustain the Steady State model (Chapter 5). Hawking’s thesis, by contrast, showed that cosmic expansion is inconsistent with the Hoyle–Narlikar theory, based on arguments from classical field theory.
Inspired by Penrose's singularity theorem for black holes \cite{Penrose64}, the final part of Hawking’s thesis focused
on the inevitability of a Big Bang singularity, assuming the validity of general relativity. With the
growing acceptance of the Big Bang theory, it is easy to understand the conclusion established in Parker's first published paper \cite{Parker68}: 

\begin{quote}
    {\it The particle creation in the expanding universe at the present time is quite negligible. However, for the early stages of a Friedmann expansion it may well be of great cosmological significance, especially since it seems inescapable if
one accepts quantum field theory and general relativity. In considering the large amount of particle creation taking place in the early stages of an expansion, it is necessary to take into account
the reaction of the matter created back on the gravitational field. Furthermore, it may be necessary
to consider the effects of the quantization of the gravitational field. Therefore, no conclusive
quantitative result can yet be reported here concerning the primeval creation.}  
\end{quote}


\section{The years after the thesis: 1966-1973}

A second important example in which a time-dependent gravitational field arises naturally in general relativity is gravitational collapse, which reaches its most extreme form when ending in a black hole. It is therefore not surprising that Parker was drawn to this problem. In his own words \cite{Parker17}:

\begin{quote}
{\it Before leaving Harvard, I suggested to my advisor that I believed it would be interesting to calculate the particle creation by the gravitational field of matter collapsing to form a black hole. He suggested that I write to John Wheeler at Princeton when applying for a postdoc position. I did that but the reply said that no postdoc position was available, but if I had funds of my own, I could come to Princeton to work on the project. I also mentioned the idea to Bryce DeWitt when I was an instructor at UNC. He agreed that the particle creation should occur, but did not express an interest in working on such a project with me.} 
\end{quote}

\subsection{United States}

Parker spent two academic years (1966–1968) at the University of North Carolina, at the Institute of Field Physics in Chapel Hill directed by  DeWitt. It is understandable that DeWitt showed little interest in Parker’s proposal at the time. His focus then remained on the development of a covariant approach to perturbative quantum gravity and Yang-Mills theory. Ironically, a decade later, DeWitt wrote an authoritative review on quantum field theory in curved spacetime and particle production  \cite{DeWitt75}—the very subject first opened by Parker’s thesis.

In 1970, Remo Ruffini (assistant professor in the group of John Wheeler) and Steven Fulling, a Princeton graduate student supervised by Arthur Wightman, visited Parker in Milwaukee. Parker was invited to visit Princeton, using his NSF grant, in the academic year 1971-1972.\footnote{On November 9, 1971, Parker gave a seminar at Princeton entitled Cosmological Particle Production \cite{Wheelerfiles}.} His grant proposal included the idea of calculating the particle creation that would occur when matter collapsed to form a black hole.   Parker was also invited by Wightman to be the second reader of  Fulling’s
thesis.  Interest in Parker's thesis and its implications was gradually growing in the United States.\footnote{In addition to Fulling, the list of prominent contributors to gravitational particle creation and quantum field theory in curved spacetime typically includes J. Bekenstein, L. Ford, B.-L. Hu, W. Unruh, and R. Wald—each of whom was a Princeton graduate student in 1971–72.}
 
In his Ph.D. thesis \cite{Fulling72, Fulling73},  Fulling demonstrated that the superposition of creation and annihilation operators—similar to (\ref{Bogo}) and arising in an expanding universe—also appears in the Rindler coordinate system of uniformly accelerated observers restricted to a wedge of Minkowski space. To account for this situation, one needs to slightly generalize (\ref{Bogo}) to Bogoliubov transformations that involve a sum over different momenta, as in
\begin{equation}\label{Bogo2}
B_{\vec{k}}=\sum_{{\vec k^{\prime}}}\left[\alpha_{\vec k, \vec k^{\prime}} A_{\vec{k}^{\prime}}+\beta_{\vec k, \vec k^{\prime}} A_{-\vec{k}^{\prime}}^{\dagger}\right]
\ . \end{equation}
  
\subsection{Moscow and Cambridge}
 
The reception of Parker’s ideas was even swifter outside the United States. Among the first to recognize their physical significance were Yakov B. Zeldovich’s group in Moscow and the relativity group in Cambridge.

Zeldovich’s team quickly launched a research program to investigate the consequences of gravitational particle production \cite{Zeldovich70},
with particular emphasis on anisotropic cosmologies \cite{Zeldovich-Starobinsky}. They were also intrigued by the implications for rotating black holes. The underlying concept echoed Schrödinger’s earlier work on the backscattering of scalar waves in an isotropically expanding universe. Zeldovich \cite{Zeldovich71} and Starobinsky \cite{Starobinsky73} realized that a rotating object could amplify scalar waves. They interpreted this phenomenon as the classical analogue of a quantum process of stimulated emission, which necessarily presupposes spontaneous emission. In this context, the time dependence of gravitational collapse plays no role; the effect arises solely from rotation, and therefore no analogous particle-creation mechanism exists for non-rotating black holes.\footnote{On the U.S. side, C. Misner \cite{Misner72}, W. Unruh \cite{Unruh74}, L. Ford \cite{Ford75}, and others were independently examining rotation as the source of quantum black hole radiation—an effect now recognized as quantum superradiance. For a review on superradiance, see \cite{Brito20}.}

One of the earliest acknowledgments from Cambridge appeared in a 1970 paper by  Hawking \cite{Hawking70}, which—although focused on classical general relativity—highlighted the importance of the frequency-mixing mechanism and the associated linear (Bogoliubov) transformation of creation and annihilation operators that Parker had introduced and developed in his thesis. As Gary Gibbons later remarked \cite{MJC07}:

\begin{quote}
 {\it Around 1972, I became interested in the processes that take place near a black hole. I was working with Stephen Hawking. We both studied professor Parker's papers. Those papers certainly influenced the calculations we made.}
 \end{quote}
 
 \section{Consequences and extensions: the lasting impact of Parker's thesis} \label{section5}

 \subsection{Black hole radiation}

Parker’s mechanism for cosmological particle creation played a pivotal role in Hawking’s groundbreaking work on black hole radiation. A central feature of Hawking’s derivation was his treatment of black hole formation as a time-dependent process—an approach that closely paralleled Parker’s analysis of a dynamically evolving universe. . 

In the case of Schwarzschild black holes, and also for the field model (\ref{action}), the analogue of the ratio (\ref{beta/alpha}) obtained at late retarded times, when the black hole has settled down to a stationary configuration,  is\footnote{To properly evaluate late-time particle creation, Hawking replaced plane wave modes with wave packets. For the sake of simplicity, we won't go into those details here. We also omit discussing the role of gray-body factors.}$^{,}$\footnote{This situation requires the use of non-diagonal Bogoliubov transformations, such as (\ref{Bogo2}). For  simplicity, we  omit the angular quantum numbers $l, m$.} \cite{Hawking75} 
\begin{equation}\label{beta/alphaHawking}
\left|\frac{\beta_{\omega \omega'}}{\alpha_{\omega \omega'}}\right|^2= e^{-4\pi M\omega} \ . 
\end{equation}
The analogue of the mean number of created particles (\ref{number}) yields a Planck distribution of thermal radiation at the (Hawking) temperature\footnote{It was also shown independently by Parker \cite{Parker75}, Wald \cite{Wald75}, and Hawking \cite{Hawking76} that  the full probability
distribution of the created particles was exactly thermal, not just the mean number distribution mentioned above. This probability distribution was derived using methods analogous to those employed in obtaining the particle creation probability distribution in an expanding universe, as discussed in the paragraph containing Eq. (\ref{beta/alpha}). Further details on black hole emission, including its dependence on particle species and the associated gray-body factors, were derived in Ref. \cite{Page76}.} 
 \begin{equation}
 T_H= \frac{\hbar c^3}{8\pi G M k_B} \ . 
 \end{equation}

This astonishing result\footnote{The result reinforced the deep connection between black holes and thermodynamics, originally proposed by J. Bekenstein \cite{Bekenstein73}, who argued that entropy is proportional to the area of the event horizon.} was first submitted as a letter to Nature in January 1974. In February, Hawking presented his findings at the First Oxford Symposium on Quantum Gravity, held at Rutherford Laboratory. Around the same time, a Cambridge preprint containing a more detailed version of the calculations and results was circulated. This version was submitted to Communications in Mathematical Physics (CMP) in March, but the editors misplaced the manuscript, and Hawking had to resubmit it in April 1975. The paper was finally published in CMP in August 1975.

Another interesting
curiosity concerns the written version of Hawking’s contribution at the Oxford Symposium, which appeared in the symposium proceedings \cite{Proceedings75} also in August 1975. It was essentially an amended version of the article submitted to CMP. The only difference was the inclusion of a citation to Parker’s paper \cite{Parker69}, added in the introduction at the end of the following paragraph: 
 
 \begin{quote}
{\it This means that the initial vacuum state $|0_1\rangle$, the state that satisfied $a_{1i}|0_1\rangle=0$ for each initial annihilation operator $a_{1i}$, will not be the same as the final vacuum state $|0_3\rangle$ i.e., $a_{3i}|0_1\rangle \neq 0$. One can interpret this as implying that the time-dependent metric or gravitational field as caused the creation of a certain number of particles of the scalar field.$^{(31)}$ }  
 \end{quote}
[citation (31) corresponds to \cite{Parker69}]. It should be noted that Hawking revised his original preprint to acknowledge Parker’s significant influence on his work. However, the omission of this citation in the final published version of Hawking’s  paper in CMP obscured the connection between Parker’s earlier contributions and the approach Hawking employed.

\subsection{Fulling-Davies-Unruh effect}

Motivated by Hawking’s thermal result, Paul Davies calculated the mean particle number from the Bogoliubov coefficients previously obtained by  Fulling for the Rindler coordinate system corresponding to a uniformly accelerated observer in Minkowski space \cite{Fulling72, Fulling73}. This calculation also yielded a thermal spectrum, with temperature \cite{Davies75}
\begin{equation}
T= \frac{\hbar a}{2\pi c k_B} \ , 
\end{equation}
where $a$ is the acceleration. 
The physical significance of this result was later deepened by William G. Unruh, who demonstrated that a uniformly accelerated detector in Minkowski space perceives the standard Minkowski vacuum as a thermal bath of radiation \cite{Unruh76}.\footnote{For a comprehensive review, see \cite{Crispino08}.}
A direct connection to Hawking’s result arises upon identifying the acceleration $a$
 with the surface gravity $\kappa$ of the Schwarzschild black hole horizon
 $\kappa= \frac{c^4}{4 G M} $. 

Further insight into the link between these developments and the framework established in Parker’s thesis can be found in the account provided by  Unruh \cite{Unruh25}:

\begin{quote}
{\it Key people whose research really influenced me:
Steve Fulling – gave informal lectures on quantum field theory,
 worked on quantum field theory
in curved spacetime (strongly influenced by Leonard Parker’s work on quantum fields
in cosmological spacetimes [1]) and worked on quantization of quantum fields in
Minkowski spacetime and in “Rindler” spacetime. }
\end{quote}

\subsection{Creation of gravitons}

One of the main results of Parker’s thesis was the demonstration that no particle creation occurs for massless fields satisfying conformally invariant equations with spins $0, 1/2, 1$, and $2$. In contrast, as also proved in Parker's thesis, massless particles governed by the minimally coupled scalar field equation are produced from the initial vacuum state as a consequence of cosmic expansion.

In a Friedmann–Lemaître–Robertson–Walker (FLRW) spacetime, Lifshitz (1946) showed that, in a particular gauge, each of the two polarization components of a linearized gravitational wave satisfies the same equation as a massless, minimally coupled scalar field. This immediately implies that gravitons of each polarization are created in precisely the same manner as the minimally coupled spin-0 particles predicted in Parker’s thesis. This correspondence was first discussed  by Leonid Grishchuk \cite{Grishchuk74}, who derived an expression for graviton creation in a universe whose scale factor 
$a(t)$ evolves as a power law in time. A more detailed and comprehensive analysis was later presented in Ref. \cite{FordParker77}.

Furthermore, in an exponentially expanding inflationary universe, well-known analytic solutions exist for the equations governing a massless, minimally coupled scalar field.  They form the foundation for the inflationary predictions of linearized gravitational waves and the associated graviton production in the early universe, as well as for the scalar gravitational perturbations responsible for the temperature anisotropies observed in the cosmic microwave background.

 \subsection{Backreaction effects in cosmology}

 Let us go back to the  text quoted at the end of section 3, in which Parker emphasized  the strong effects that particle creation would have on
the early expansion of the universe, so the reaction-back would have to be taken into account. The first calculation of this kind was carried out by Parker and Fulling  in \cite{Parker-Fulling73}. Using the semi-classical Einstein equations, they demonstrated that a cosmological bounce could arise from the process of particle creation, thereby avoiding the cosmological singularity. Their result also showed that the classical energy conditions can be violated through particle production in quantum field theory within an FLRW universe. 

In their analysis, they avoided addressing the problem of renormalizing the stress–energy tensor, which was later examined in detail in subsequent works \cite{Parker-Fulling74}. This was followed by more in-depth studies of backreaction effects \cite{Hu-Parker78, Fischetti-Hartle-Hu}, including the role of conformal anomalies and the analysis of the damping of anisotropies and inhomogeneities.

\subsection{Inflation}

Shortly after the proposal of the inflationary universe, the creation of scalar perturbations was analyzed in detail \cite{Mukhanov-Chibisov81, Hawking82, Guth-Pi82, Starobinsky82, Bardeen-Steinhardt83}. It led to the prediction that small density perturbations would be generated in the expanding universe with an almost scale-free spectrum. Cosmological particle creation also provides the underlying mechanism driving these primordial perturbations. These perturbations seeded the tiny fluctuations in temperature observed in the cosmic microwave background, as first observed by the COBE satellite and confirmed by many other experiments, including the PLANCK satellite. It also helps explain how matter clustered to form galaxies, galactic clusters, and ultimately the large-scale structure of the universe.

\subsection{Quantum field theory in curved spacetime}

The phenomenon of cosmological particle creation, first discovered and thoroughly analyzed in Parker’s pioneering thesis, together with its subsequent extensions and refinements, opened up an entirely new domain in theoretical physics: quantum field theory in curved spacetime and the broader framework of semiclassical gravity. Originating in the 1960s and becoming firmly established throughout the 1970s and 1980s—with major emphasis on understanding the renormalization of quantum fields in curved backgrounds—this discipline was steadily developed and disseminated through numerous  review articles and monographs \cite{DeWitt75, Parker77, Isham77, Parker79, Gibbons79, Birrell-Davies, Fulling89, Wald94, GMM94, Frolov-Novikov98, FN05, Mukhanov-Winitzki07, Crispino08, Parker-Toms09, Hollands-Wald15, Hu-Verdaguer20, Ford21, Buchbinder-Shapiro22}. Its central objective has been to clarify how quantum effects manifest themselves in the presence of gravitation, especially in regimes where spacetime curvature cannot be neglected but a full theory of quantum gravity is not yet required.
A further step in this direction  is provided by the framework of stochastic gravity \cite{Hu-Verdaguer08, Hu-Verdaguer20}, which extends semiclassical gravity by incorporating the fluctuations of the quantum stress–energy tensor beyond its mean expectation value. This yields a richer description of backreaction and provides a theoretical bridge between semiclassical and fully quantum gravitational regimes. In parallel, gravitational particle creation has also played a fundamental role
in boosting the algebraic approach to the theory of quantized fields in curved spacetime \cite{Fulling72, Ashtekar-Magnon75, Kay85, Hollands-Wald15}. Rather than relying on specific  choices of preferred vacuum states, the   algebraic approach constructs quantum fields as nets of local algebras associated with spacetime regions, emphasizing locality, covariance, and general geometric structure.

\section{One step backward, two steps forward: the legacy of Parker’s 1966 thesis} \label{section6}

In this paper, we aimed to highlight two key features of Parker's Thesis: its originality within its historical context and the relevance of its results to the overall understanding of the intersection between quantum and gravity.

Indeed, in 1962, Parker’s PhD project stood apart from the mainstream. At that time, most efforts were directed toward developing a  theory of quantum gravity, exemplified by the deep work of DeWitt and others. However, 
this reorientation of the research -- combining gravity and quantum theory in the most conservative framework (one step backward) -- proved to be remarkably fruitful. To illustrate this perspective, in his lecture at the Oxford Symposium in February 1974 \cite{Isham74}, Chris J. Isham remarked: 

\begin{quote}

{\it From a theoretical point of view a thorough understanding of this simple model}  [he refers to the model (\ref{action})] {\it would seem a natural prerequisite to attempting to quantise the metric tensor itself. It is therefore perhaps surprising that, whereas considerable amount of effort has been expended over that last twenty five years on the full quantum gravity theory, only a relatively small amount of work has appeared dealing with this simplified problem. }
\end{quote}

But Parker's results shouldn't be taken as merely consistency checks between quantum field theory and general relativity, but as the first step to a major independent research field, revealing many hidden insights and unexpected  (two steps forward). As we have noted in section 5, the spreading of research programs in black holes, particle detectors, or cosmology has still maintained a continuous (although sometimes not linear) activity until today.  Without attempting to be exhaustive, we note that cosmological particle creation continues to play a significant role as a potential explanation for the properties of dark matter, as emphasized in \cite{Kolb24}. On the mathematical side, the concept of adiabatic states introduced in Parker's original work—central to modern theoretical cosmology—has been naturally extended to the more general framework of Hadamard states, which play a fundamental role in the foundational structure of quantum field theory in curved spacetime \cite{Wald94}. 
 Gravitational particle creation is still a cornerstone of quantum gravity. In this regard, DeWitt noted in a retrospective paper \cite{DeWitt09} that the modern history of quantum gravity begins with the discovery of black hole radiation.\footnote{Recovering the results on black-hole particle creation—whether in the form of black-hole emission or through the Bekenstein–Hawking entropy formula—has provided a profound stimulus to research in quantum gravity, inspiring advances in both string theory and loop quantum gravity.}
Squeezed quantum states arising from gravitational particle creation—exemplified in a particularly clear way by the Fulling–Davies–Unruh effect—have played a major role in shaping recent research directions on entanglement in quantum field theory \cite{Witten18}.
Equally noteworthy is the highly active field of analogue gravity \cite{Barcelo05}, initiated with the aim of detecting laboratory analogues of gravitational particle creation. 

In summary, Parker’s thesis marked the first glimpse that the marriage of quantum  and gravity is not only conceivable but also a source of deep and surprising discoveries. The originality of his work lies not in proposing a completely new model or framework, but in a careful examination of the fundamentals of two pillars of modern physics: quantum field theory and general relativity. It is tempting to say that Parker’s vision was somewhat inspired by Einstein’s 1905 paper on special relativity, where a careful reanalysis of electromagnetism—the best-understood theory of the time—led to revolutionary insights. Likewise, Parker’s meticulous study of quantum fields in time-dependent gravitational backgrounds gave rise to a series of important and far-reaching discoveries.
In this sense, we truly believe in the enduring relevance of Parker's thesis, not only for the originality of its content side but also for the  distinctiveness of its  conceptual approach to theoretical physics.\\

{\bf Acknowledgments.} We are all very grateful to Gloria Parker for her help in facilitating the retyping and new edition of Leonard Parker’s thesis. This work has been supported by Project No. PID2023-149560NB- C21 funded by MCIU /AEI/10.13039/501100011033 / FEDER, UE. The work of SP was funded by the Deutsche Forschungsgemeinschaft (DFG, German Research Foundation) under Germany’s Excellence Strategy – EXC 2094 – 390783311.

\end{document}